\documentclass[useAMS,usenatbib]{mn2e}

\usepackage{mathptmx}
\usepackage{amssymb}
\usepackage[fleqn]{amsmath}
\usepackage{url}
\usepackage{graphicx}
\usepackage{float}
\usepackage{enumitem}
\usepackage{txfonts}
\usepackage{bibentry} 
\usepackage{etoolbox} 

\usepackage{xspace}

\newcommand{\SimpleX}{\textsc{simplex}\xspace}
\newcommand{\MdotA}{\ifmmode{\dot{M}_{\eta_{\mathrm{A}}}}\else{$\dot{M}_{\eta_{\mathrm{A}}}$}\fi\xspace}
\newcommand{\MdotB}{\ifmmode{\dot{M}_{\eta_{\mathrm{B}}}}\else{$\dot{M}_{\eta_{\mathrm{B}}}$}\fi\xspace}
\newcommand{\etaA}{$\eta_{\mathrm{A}}$\xspace}
\newcommand{\etaB}{$\eta_{\mathrm{B}}$\xspace}
\newcommand{\ec}{$\eta$~Car\xspace}
\makeatletter
\newcommand\bibstyle@comma{\bibpunct{(}{)}{,}{a}{}{,}}
\newcommand\bibstyle@semicolon{\bibpunct{(}{)}{;}{a}{}{,}}
\makeatother
\pretocmd\citet{\citestyle{comma}}\relax\relax
\pretocmd\citep{\citestyle{semicolon}}\relax\relax

\newcommand{\Ms}{\ifmmode{~\mathrm{M}_{\odot}}\else{$\mathrm{M}_{\odot}$}\fi\xspace}
\newcommand{\Msy}{\ifmmode{\Ms \per{yr}{-1}}\else {$\Ms \per{yr}{-1}$}\fi\xspace}
\newcommand{\Ls}{\ifmmode{~\mathrm{L}_{\odot}}\else{$\mathrm{L}_{\odot}$}\fi\xspace}
\newcommand{\kms}{\ifmmode{~\mathrm{km}\per{s}{-1}}\else {$\mathrm{km}\per{s}{-1}$}\fi\xspace}
\newcommand{\Rs}{\ifmmode{~\mathrm{R}_{\odot}}\else{$\mathrm{R}_{\odot}$}\fi\xspace}
\newcommand{\per}[2]{\ifmmode{\mathrm{\,#1}^{#2}}\else {$\mathrm{\,#1}^{#2}$}\fi\xspace}

\newcommand{\ten}[1]{\ifmmode{10^{#1}}\else{$10^{#1}$}\fi\xspace}
\newcommand{\sci}[2]{\ifmmode{#1 \times 10^{#2}}\else{$#1 \times 10^{#2}$}\fi\xspace}

\newcommand{\HI}{\ifmmode{\mathrm{H\,I}}\else{H\textsc{$\,$i}}\fi\xspace}
\newcommand{\HII}{\ifmmode{\mathrm{H\,II}}\else{H\textsc{$\,$ii}}\fi\xspace}
\newcommand{\HeI}{\ifmmode{\mathrm{He\,I}}\else{He\textsc{$\,$i}}\fi\xspace}
\newcommand{\HeII}{\ifmmode{\mathrm{He\,II}}\else{He\textsc{$\,$ii}}\fi\xspace}
\newcommand{\HeIII}{\ifmmode{\mathrm{He\,III}}\else{He\textsc{$\,$iii}}\fi\xspace}
\newcommand{\FeIII}{\ifmmode{[\mathrm{Fe\,III}]}\else{[Fe\textsc{$\,$iii}]}\fi\xspace}
\newcommand{\FeII}{\ifmmode{[\mathrm{Fe\,II}]}\else{[Fe\textsc{$\,$ii}]}\fi\xspace}
\newcommand{\NiII}{\ifmmode{[\mathrm{Ni\,II}]}\else{[Ni\textsc{$\,$ii}]}\fi\xspace}
\newcommand{\nuHI}{\ifmmode{\nu_{\HI}}\else{$\nu_{\HI}$}\fi\xspace}
\newcommand{\nuHeI}{\ifmmode{\nu_{\HeI}}\else{$\nu_{\HeI}$}\fi\xspace}
\newcommand{\nuHeII}{\ifmmode{\nu_{\HeII}}\else{$\nu_{\HeII}$}\fi\xspace}

\newcommand{\ion}[2]{\ifmmode{\mathrm{#1}^{#2}}\else{#1$^{#2}$}\fi\xspace}
\newcommand{\eal}[2]{\ifmmode{\mathrm{#1\,#2}}\else{#1\textsc{$\,$\lowercase{#2}}}\fi\xspace}
\newcommand{\nuion}[2]{\ifmmode{\nu_{\ion{#1}{#2}}}\else{$\nu_{\ion{#1}{#2}}$}\fi\xspace}

\defcitealias{Madura_etA_2013}{M13}
\defcitealias{Clementel_etA_2014a}{C14a}
\defcitealias{Clementel_etA_2014b}{C14b}

\usepackage{color}

\usepackage{ifdraft}

\ifdraft{\newcommand{\rmf}{\comment}} {\newcommand{\rmf}{}}

\title[\ec He ionization structure at periastron]{3D radiative transfer simulations of Eta Carinae's inner colliding winds -- II: Ionization structure of helium at periastron}

\author[Clementel et al.]{N. Clementel$^{1}$\thanks{E-mail: clementel@strw.leidenuniv.nl}, T.~I. Madura$^{2}$, C.~J.~H. Kruip$^{1}$ and J.-P. Paardekooper$^{3,4}$\\
$^{1}$Leiden Observatory, Leiden University, PO Box 9513, 2300 RA Leiden, the Netherlands\\
$^{2}$Astrophysics Science Division, Code 667, NASA Goddard Space Flight Center, Greenbelt, MD 20771, USA\\
$^{3}$Universit\"at Heidelberg, Zentrum f\"ur Astronomie, Institut f\"ur Theoretische Astrophysik, Alber--Ueberle--Str. 2, 69120 Heidelberg, Germany\\
$^{4}$Max Planck Institute for Extraterrestrial Physics, PO Box 1312, Giessenbachstr., D--85741 Garching, Germany\\
}

\begin{document}

\date{Accepted ***. Received ***; in original form ***}

\pagerange{\pageref{firstpage}--\pageref{lastpage}} \pubyear{2014}

\maketitle

\label{firstpage}

\begin{abstract}

Spectral observations of the massive colliding wind binary Eta~Carinae show phase-dependent variations, in intensity and velocity, of numerous helium emission and absorption lines throughout the entire 5.54-year orbit. Approaching periastron, the 3D structure of the wind--wind interaction region (WWIR) gets highly distorted due to the eccentric ($e \sim 0.9$) binary orbit. The secondary star (\etaB) at these phases is located deep within the primary's dense wind photosphere. The combination of these effects is thought to be the cause of the particularly interesting features observed in the helium lines at periastron. We perform 3D radiative transfer simulations of \ec's interacting winds at periastron. Using the \SimpleX radiative transfer algorithm, we post-process output from 3D smoothed particle hydrodynamic simulations of the inner 150~au of the \ec system for two different primary star mass-loss rates (\MdotA). Using previous results from simulations at apastron as a guide for the initial conditions, we compute 3D helium ionization maps. We find that, for higher \MdotA, \etaB\ \ion{He}{0+}-ionizing photons are not able to penetrate into the pre-shock primary wind. \ion{He}{+} due to \etaB is only present in a thin layer along the leading arm of the WWIR and in a small region close to the stars. Lowering \MdotA allows \etaB's ionizing photons to reach the expanding unshocked secondary wind on the apastron side of the system, and create a low fraction of \ion{He}{+} in the pre-shock primary wind. With apastron on our side of the system, our results are qualitatively consistent with the observed variations in strength and radial velocity of \ec's helium emission and absorption lines, which helps better constrain the regions where these lines arise.

\end{abstract}



\section{Introduction}\label{intro}

The massive colliding wind binary Eta Carinae (\ec) is one of the most luminous stellar objects in the Galaxy ($L_{\mathrm{Total}} \gtrsim$~$\sci{5}{6} \Ls$). During its highly eccentric ($e \sim 0.9$), 5.54~yr orbit, the slow but dense wind of the primary star \etaA ($v_{\infty}$~$\approx$~420~\kms, $\MdotA$~$\approx$~$\sci{8.5}{-4} \Msy$; \citealt{Hillier_etA_2001, Groh_etA_2012a}) collides with the faster ($v_{\infty} \approx 3000 \kms$; \citealt{Pittard_Corcoran_2002, Parkin_etA_2009}), but less dense ($\MdotB$~$\approx$~$\sci{1.4}{-5} \Msy$) wind of the secondary star \etaB. Every periastron, orbital motion and the binary's large eccentricity highly distort the stellar winds and wind-wind interaction region (WWIR), causing \etaB to become deeply embedded within the denser \etaA wind \citep{Okazaki_etA_2008, Madura_2010, Parkin_etA_2011, Madura_etA_2012, Madura_etA_2013}. At periastron, the dense distorted \etaA wind and WWIR trap the ionizing radiation from \etaB that is responsible for the formation of numerous high-ionization (ionization potential $\gtrsim$~13.6~eV) emission and absorption lines observed during the broad part of the orbit around apastron \citep{Verner_etA_2005, Damineli_etA_2008b, Gull_etA_2009, Madura_2010, Gull_etA_2011, Madura_Groh_2012, Madura_etA_2012, Madura_etA_2013, Teodoro_etA_2013}.

In a recent paper, \citet[][hereafter C14b]{Clementel_etA_2014b} presented results from 3D radiative transfer (RT) simulations focusing on the ionization structure of helium, due to \etaB's ionizing radiation, in the inner $\sim$~155~au of the \ec system at an orbital phase of apastron. As summarized in \citetalias{Clementel_etA_2014b}, helium spectral features provide important information on both the geometry and physical properties of the \ec binary and the individual stars. While various helium features are present throughout the entire orbit, they show their most interesting behavior during periastron passage \citep{Nielsen_etA_2007, Damineli_etA_2008b, Teodoro_etA_2012}.

The goal of this paper is to extend the work of \citetalias{Clementel_etA_2014b} and analyze the effects of \etaB's ionizing radiation during the spectroscopic low state at periastron. As in \citetalias{Clementel_etA_2014b}, we focus on the inner $\sim$~155~au of the \ec system and apply radiative transfer post-processing simulations to 3D smoothed particle hydrodynamics (SPH) simulations of \ec's colliding winds \citep[][hereafter M13]{Madura_etA_2013} to compute 3D maps of the ionization structure of He. This work sets the stage for future efforts to compute synthetic He line profiles for direct comparison with available observations. The goal is to constrain where and how the observed broad He emission and absorption lines form during the periastron event. We further aim to explain the temporal behavior of the broad He lines as the \ec system moves from apastron through periastron. Again, we note that we focus solely on interpreting the broad emission and absorption features of He that arise in the stellar winds and WWIRs, and not the narrower ($\lesssim$~50~\kms) features that form in the Weigelt blobs and other dense, slow-moving near-equatorial circumstellar ejecta \citep{Weigelt_Ebersberger_1986, Damineli_etA_2008b}.

This paper is organized as follows. In Section~\ref{sec:Method}, we summarize our numerical approach. Section~\ref{sec:Results} describes the results. A discussion of the results and their implications is in Section~\ref{sec:Discussion}. Section~\ref{sec:Summary} summarizes our conclusions and outlines the direction of future work.

\section{Methods}\label{sec:Method}

The numerical approaches in this paper, both for the SPH and RT simulations, are identical to those in \citetalias{Clementel_etA_2014b}. As in previous works, we use the \SimpleX algorithm \citep{Ritzerveld_Icke_2006, Ritzerveld_2007, Paardekooper_etA_2010, Kruip_etA_2010, Kruip_2011} to post-process 3D SPH simulation output \citepalias{Madura_etA_2013}. For further details, we refer the reader to \citet{Clementel_etA_2014a}, \citetalias{Clementel_etA_2014b}, and references therein. We also retain the naming convention used in \citetalias{Madura_etA_2013}, \citet{Clementel_etA_2014a}, and \citetalias{Clementel_etA_2014b} when referring to the SPH and \SimpleX simulations in this paper, namely, Case~A ($\MdotA = \sci{8.5}{-4}$~\Msy), Case~B ($\MdotA = \sci{4.8}{-4}$~\Msy), and Case~C ($\MdotA = \sci{2.4}{-4}$~\Msy). In the following, we briefly describe only the key aspects and differences that are directly relevant to this work.

Due to the high eccentricity of the \ec system ($e \sim 0.9$), as the stars approach periastron, both the 3D structure and dynamical evolution of the individual stellar winds and WWIR change drastically, compared to the majority of the orbital period. The stellar separation drops to $\sim$~1.5~au and orbital speeds increase greatly, becoming comparable to the wind speed of \etaA (see Section~\ref{sec:Results} for details). The rapid orbital motion during periastron thus leads to major changes in the structure of the winds and WWIR on relatively short time-scales of the order of a few days (\citealt{Okazaki_etA_2008, Parkin_etA_2011, Madura_etA_2012}; \citetalias{Madura_etA_2013}).

To properly resolve this complex situation, we use two different SPH simulation domain sizes from \citetalias{Madura_etA_2013}. The larger domain is the same as the one used in \citetalias{Clementel_etA_2014b} ($r = 10\,a = 155$~au, where $a$~$=$~15.45~au is the length of the orbital semimajor axis), and allows for a direct comparison of the results in this paper with those at apastron in \citetalias{Clementel_etA_2014b}. The smaller domain simulations have a computational domain radius $r = 1.5\,a = 23$~au and eight times the number of SPH particles used in the $r = 10\,a$ simulations, leading to a roughly factor of two improvement in the overall spatial resolution compared to the $r = 10\,a$ simulations. The high-resolution simulations focus only on phases around periastron ($\phi = 0.97$--1.03, where by convention, $\phi = t/2024 = 0, 1, 2, ...$, is defined as periastron, and $\phi = 0.5, 1.5, ...$, is defined as apastron), with output written every $\Delta \phi \approx \sci{8}{-4}$, which is approximately every 1.6~days of the 2024-day orbit. These smaller, higher-resolution simulations ensure that we are adequately resolving the stellar winds and WWIR directly between the stars, as well as any possible WWIR `collapse' that may affect the escape of \etaB's ionizing radiation \citepalias{Madura_etA_2013}. The higher resolution $r = 1.5\,a$ simulations are also useful for checking the accuracy of the $r = 10\,a$ results.

We consider the ionization of hydrogen and helium atoms by both collisional- and photoionization, and assume the same abundance by number of He relative to H as \citet{Hillier_etA_2001}, $n_{\mathrm{He}} / n_{\mathrm{H}} = 0.2$. We employ a single photoionizing source located at the position of \etaB. As in \citetalias{Clementel_etA_2014b}, we consider \etaB to be an O5 giant with $\mathrm{T_{eff}} \approx 40,000$~K, we assume a total ionizing flux for H and He of \sci{3.58}{49} photons\per{s}{-1} \citep{Martins_etA_2005}, and we use three bins to sample the stellar spectrum, which we approximate with a black body. The bin edges of each frequency bin are set by the ionization energy of each species. The first bin ranges from the ionization frequency of \ion{H}{0+} (\nuion{H}{0+} = \sci{3.28}{15}~Hz) to that of \ion{He}{0+} (\nuion{He}{0+} = \sci{5.93}{15}~Hz), the second from \nuion{He}{0+} to \nuion{He}{+} (\sci{1.31}{16}~Hz), and the third from \nuion{He}{+} to a maximum frequency equal to ten times \nuion{H}{0+}. We use an effective cross-section representation to determine the correct number of absorptions within each frequency bin. We set the black body temperature to the value that produces the correct ratio $\mathrm{photons}_{\ion{H}{0+}} / \mathrm{photons}_{\ion{He}{0+}}$ (in this case $T_{\mathrm{bb}} = 49,000$~K), in accordance with the $q_{\ion{H}{0+}} / q_{\ion{He}{0+}}$ ratio in \citet{Martins_etA_2005}. $q_{\ion{He}{+}}$ is effectively zero for \etaB.

In figures displaying our \SimpleX results for density and temperature, we visualize the average of the four vertices that compose each Delaunay cell. Unfortunately, this approach leads to cell values that are difficult to interpret if the vertex values differ by several orders of magnitude. This might occur for the fractions of \ion{He}{0+}, \ion{He}{+}, and \ion{He}{2+}. In order to visualize results for these quantities in an understandable way that more truthfully represents the physics of our RT simulations, we adopt the visualization approach described in section~2.2.5 of \citetalias{Clementel_etA_2014b}. In all figures showing ionization structure, we display the minimum vertex value for the fraction of \ion{He}{0+} and the maximum vertex value for the fraction of \ion{He}{2+}. For \ion{He}{+}, we show the maximum value whenever the temperature of the gas is lower than \ten{5}~K, and the minimum otherwise (for temperatures $> \ten{5}$~K, the fraction of \ion{He}{+}, due to collisional ionization, is $\approx$~\ten{-3} or lower). For the reasons discussed in section~2.3.1 of  \citetalias{Clementel_etA_2014b} (i.e. H is already ionized in the inner $\sim$~120--150 au region centred around \etaA), we do not include plots of the ionization structure of H. We emphasize that the visualizations are merely to help guide the reader, and neither the physics nor the conclusions of our work depend on them (see \citetalias{Clementel_etA_2014b} for details).

Using the method described in section~2.3.1 of \citetalias{Clementel_etA_2014b}, we account for the inner \ion{He}{+} volume around \etaA ($r \sim$~3~au for Case~A, and $\sim$~7.5~au for Case~B) based on 1D \textsc{cmfgen} models of \ec by \citet{Hillier_etA_2001, Hillier_etA_2006}. We do not include the inner \ion{He}{2+} zone in \etaA's wind in the $r = 10\,a$ simulations since it is of negligible size ($r < 1$~au), and because \etaB produces essentially zero \ion{He}{+}-ionizing photons. Due to their smaller domain size, for added accuracy, the 23~au simulations do include the \ion{He}{2+} structure in \etaA's innermost wind ($r \sim$~0.7~au for Case~A, and $\sim$~0.8~au for Case~B).

As discussed in \citetalias{Clementel_etA_2014b}, our results at apastron appear to rule out low values of \MdotA, such as that in Case~C. This is supported by the work of \citetalias{Madura_etA_2013} and recent observations of \ec (T.~Gull and M.~Corcoran, private communication). Therefore, we focus in this paper on simulation Cases~A and B, which represent the most likely values of \etaA's current mass loss rate. Moreover, little information would be obtained by including Case~C since for such a low mass loss rate \etaA already singly-ionizes He throughout the entire computational domain in the $r = 1.5\,a$ simulations, and nearly the entire domain (out to $r \approx 120$~au) in the $r = 10\,a$ simulations.

Finally, we note that we use a standard Cartesian coordinate system with the origin located at the system centre of mass and the orbital plane set in the $xy$ plane, with the major axis along the $x$-axis. The stars orbit counterclockwise if viewed from along the $+z$-axis. In this system of reference, at periastron, \etaA is to the right and \etaB is to the left.

\subsection{Initial ionization state of the gas}

As in \citetalias{Clementel_etA_2014b}, we post-process the SPH simulation output. This approach works well for the \ec system at apastron, when the ionization state reaches an equilibrium value on a time-scale much shorter than the orbital time-scale. More importantly, at apastron, the \etaB wind cavity is an almost axisymmetric cone, allowing \etaB's ionizing photons to completely ionize the low-density secondary wind. In other words, it is reasonable at apastron to assume that knowledge of the ionization structure of the preceding SPH timestep is not required for accurate RT results.

In this paper, however, we deal with phases at periastron where, due to rapid orbital motion, the interacting wind structures evolve on a time-scale comparable to the time necessary for the RT simulations to reach a stable ionization state. Moreover, the \etaB wind cavity at periastron is strongly asymmetrical and distorted. Regions that were ionized by \etaB in an earlier snapshot may not be reachable by \etaB's photons in the next snapshot. To tackle these issues, we use a slightly different approach than that in \citetalias{Clementel_etA_2014b}.

Using collisional ionization equilibrium, the results of \citetalias{Clementel_etA_2014b}, and estimates for the recombination time of \ion{He}{+} as a function of gas density and temperature, we are able to better constrain the initial ionization state of the gas for use as an initial condition in our periastron \SimpleX simulations. We know that \ion{He}{0+} in the cooler ($T \approx \ten{4}$~K) unshocked gas in the \etaB wind on the apastron side of the system is completely photoionized by \etaB during most of the binary orbit around apastron \citepalias{Clementel_etA_2014b}. Observations with the \emph{Hubble Space Telescope}/Space Telescope Imaging Spectrograph (\emph{HST}/STIS) show that numerous high-ionization lines that require \etaB's ionizing flux are present for most of the orbit and do not start to fade until phase $\phi \approx 0.984$ (approximately one month before periastron, \citealt{Gull_etA_2009}; \citetalias{Madura_etA_2013}). Therefore, we can safely assume that \etaB is able to completely ionize \ion{He}{0+} throughout its unshocked wind at least until $\phi \approx 0.985$ (and at least for the length scales of interest here, $r \lesssim 155$~au from the stars, since the \emph{HST} observations show ionized structures extending several thousand au from the stars). After $\phi \approx 0.985$, \etaB becomes deeply embedded in the dense wind of \etaA and the WWIR becomes so distorted by orbital motion that we can no longer assume that \etaB can efficiently ionize \ion{He}{0+} throughout its entire cool unshocked wind. If the He in the low-density ($n \lesssim \ten{6}$~\per{cm}{-3}) unshocked \etaB gas is no longer photoionized by the star, then, at $T = \ten{4}$~K, \ion{He}{+} should recombine on a time-scale $t_{\mathrm{rec}} \approx  42.4$~days (or longer). Thus, around periastron, for gas with $T \approx \ten{4}$~K and $n \approx \ten{6}$~\per{cm}{-3}, the recombination time-scale for \ion{He}{+} is about an order of magnitude larger than the orbital time-scale.

Any gas that recombines very slowly compared to the orbital time-scale around periastron poses a potential problem for the accuracy of our RT simulations, mainly because we are post-processing the SPH simulation output. The SPH output contains no information about the previous ionization state of the gas before we start our \SimpleX simulations. However, any gas with a low enough recombination rate should remain ionized from one snapshot to the next, even though \etaB photons no longer reach that material\footnote{Of course, this depends on how long it has been since the supply of ionizing photons has been cut off.}. Therefore, we should preserve the ionization state of low-recombination-rate gas in the initial condition of our RT simulations.

The time difference between $\phi = 0.985$ and periastron is $\approx$~30~days. The recombination time for \ion{He}{+} gas with $T \approx \ten{4}$~K and $n \approx \ten{6}$~\per{cm}{-3} is about two weeks longer than this. Lower-density gas at $T \approx \ten{4}$~K will have even longer recombination times. If we assume that the \etaB ionizing flux becomes trapped by \etaA's dense wind and the WWIR at $\phi \approx 0.985$, then at periastron, any such low-recombination-rate gas that was initially photoionized by \etaB at $\phi \approx 0.985$ should still be ionized.

Therefore, to mimic the fact that the low-density, cool \etaB wind on the apastron side of the system should remain ionized at periastron, we set the He fractions in the \etaB wind material with $n < \ten{7}$~\per{cm}{-3} and $T < \sci{4}{4}$~K to be $\sim$~97\% \ion{He}{+}. This should generate an initial condition closer to the previous ionization state of the photoionized \etaB gas, since for gas with $n < \ten{7}$~\per{cm}{-3} and $T < \sci{4}{4}$~K, the \ion{He}{+} recombination time-scale is $\approx$~4~days or longer. The use of collisional ionization equilibrium as an initial condition for the remaining gas remains a valid approximation even around periastron, since for the pre- and post-shock \etaA wind, the high densities allow the gas to recombine almost instantaneously. Hot ($T > \sci{4}{4}$~K) gas in the post-shock \etaB wind is collisionally ionized to \ion{He}{+} or higher.

Due to the shorter dynamical time-scales around periastron, which alter the gas distribution from snapshot to snapshot, we must also run \SimpleX for a much shorter simulation time, compared to that used at apastron in \citetalias{Clementel_etA_2014b}. In our simulation testing we find that, setting the initial condition of the gas as described above, in Case~A, the ionization state reaches an equilibrium value in less than one day. In Case~B, the gas reaches ionization equilibrium in $\sim$~2--3~days. Since significant changes to the gas density and temperature distributions around periastron also occur on a time-scale of $\sim$~2--3~days, we use a total RT simulation time of 2.5~days and a simulation time-step of $\sim 5$~minutes in order to achieve an accurate RT calculation of the ionization volumes and fractions. At these time-scales, the gas distribution, even around periastron, can be assumed to be roughly constant.


\section{Results}\label{sec:Results}

\begin{figure*}
  \begin{center}
   \rmf{\includegraphics[width=174mm]{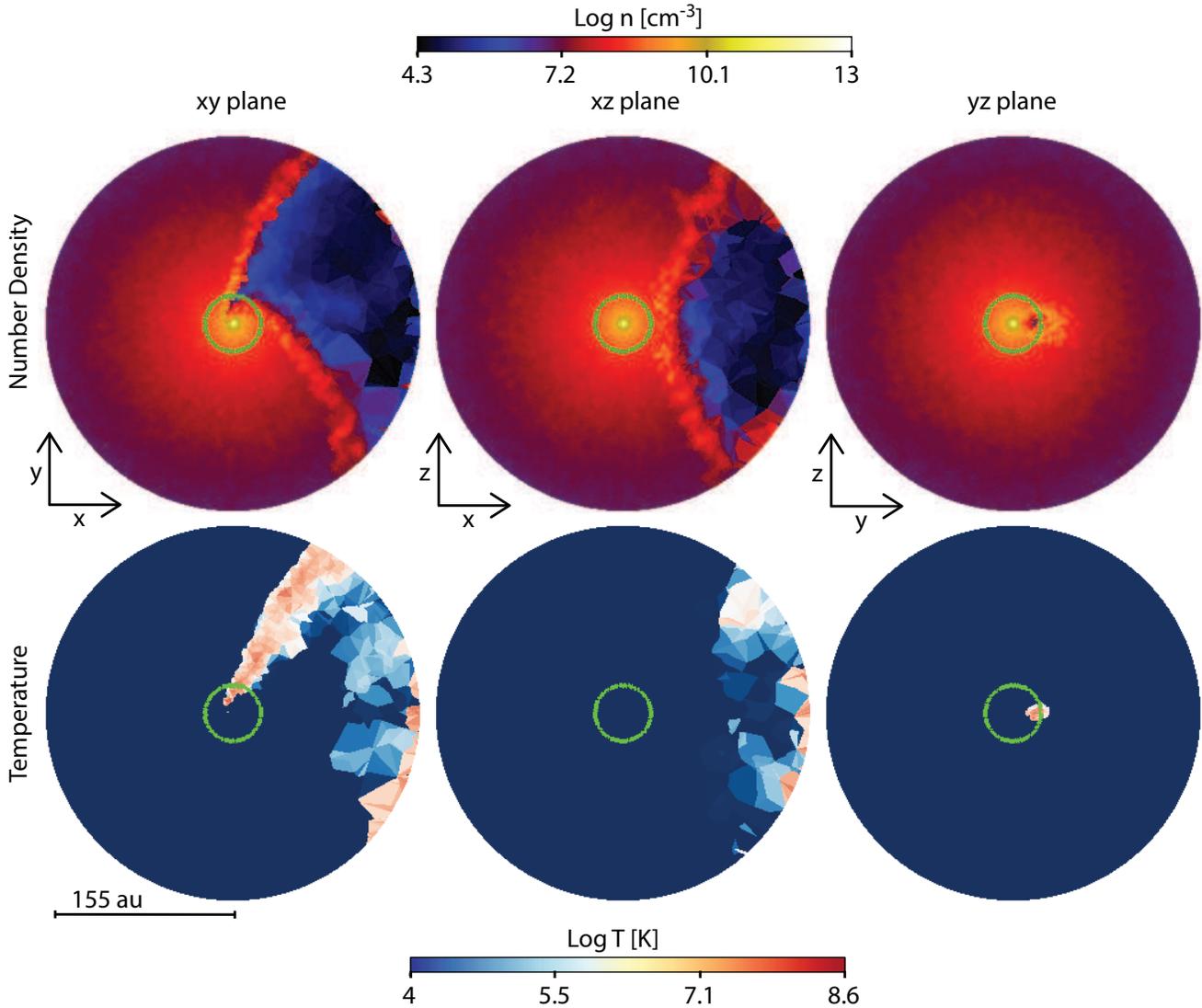}}
    \caption{Slices in the orbital $xy$ (left column), $xz$ (middle column) and $yz$ (right column) planes through the larger ($10\,a$) 3D simulation volume for the Case~A simulation of \ec at periastron. Rows show the number density (top, log scale, cgs units) and temperature (bottom, log scale, K). The length scale is shown under the bottom left panel. In the first column (i.e. the orbital plane) \etaA is to the right and \etaB is to the left. The green circle marks the boundary of the smaller ($r = 1.5a = 23$~au) domain simulations (see Fig.~\ref{fig:CaseA_small_dens_temp}).}\label{fig:CaseA_medium_dens_temp}
  \end{center}
\end{figure*}

\begin{figure*}
  \begin{center}
   \rmf{\includegraphics[width=174mm]{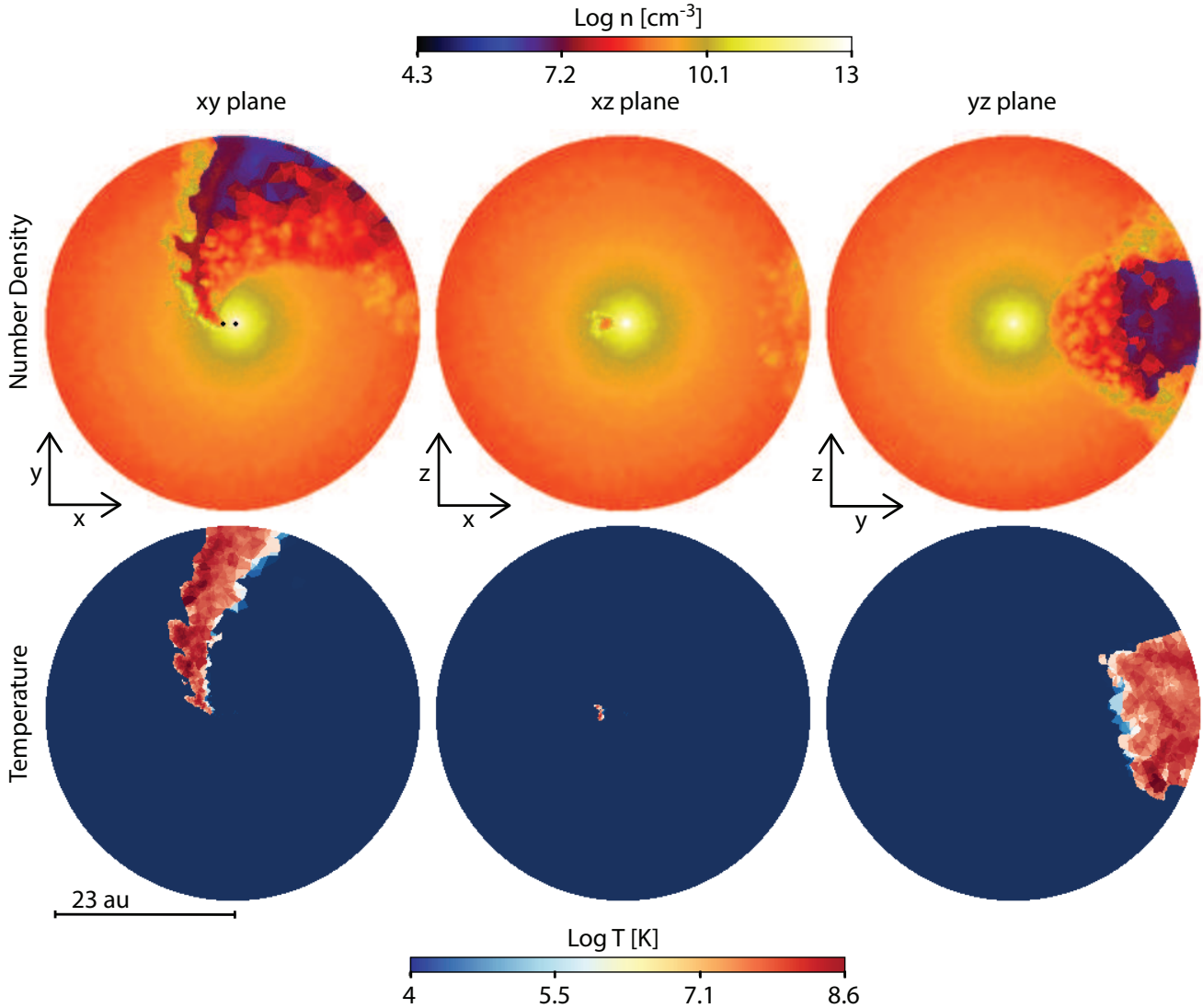}}
    \caption{Same as Fig.~\ref{fig:CaseA_medium_dens_temp}, but for the smaller (i.e. $1.5\,a$) simulation domain. The black dots in the centre of the top left panel mark the positions of the stars.}\label{fig:CaseA_small_dens_temp}
  \end{center}
\end{figure*}

\begin{figure*}
  \begin{center}
   \rmf{\includegraphics[width=174mm]{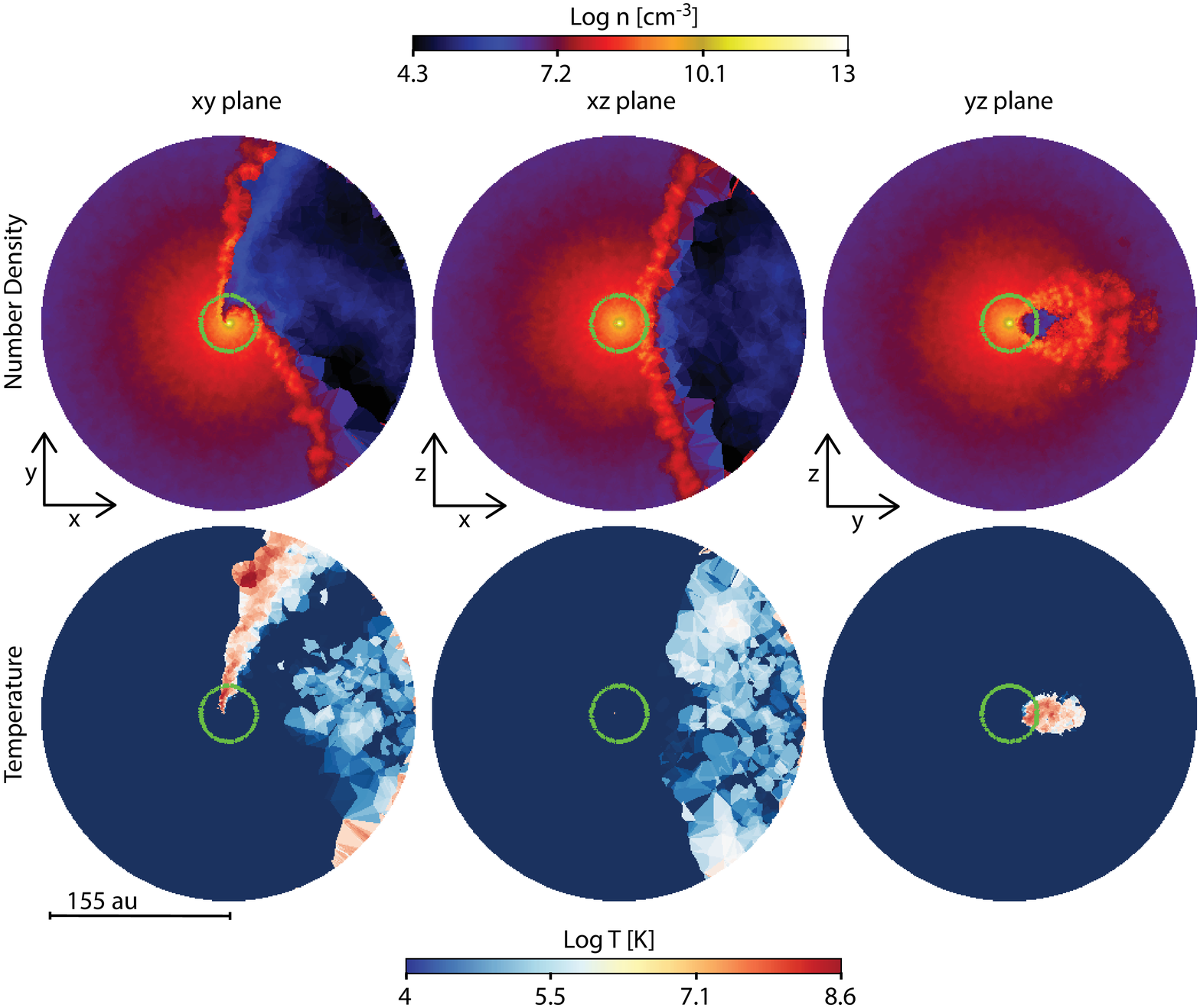}}
    \caption{Same as Fig.~\ref{fig:CaseA_medium_dens_temp}, but for the Case~B simulation.}\label{fig:CaseB_medium_dens_temp}
  \end{center}
\end{figure*}

\begin{figure*}
  \begin{center}
   \rmf{\includegraphics[width=174mm]{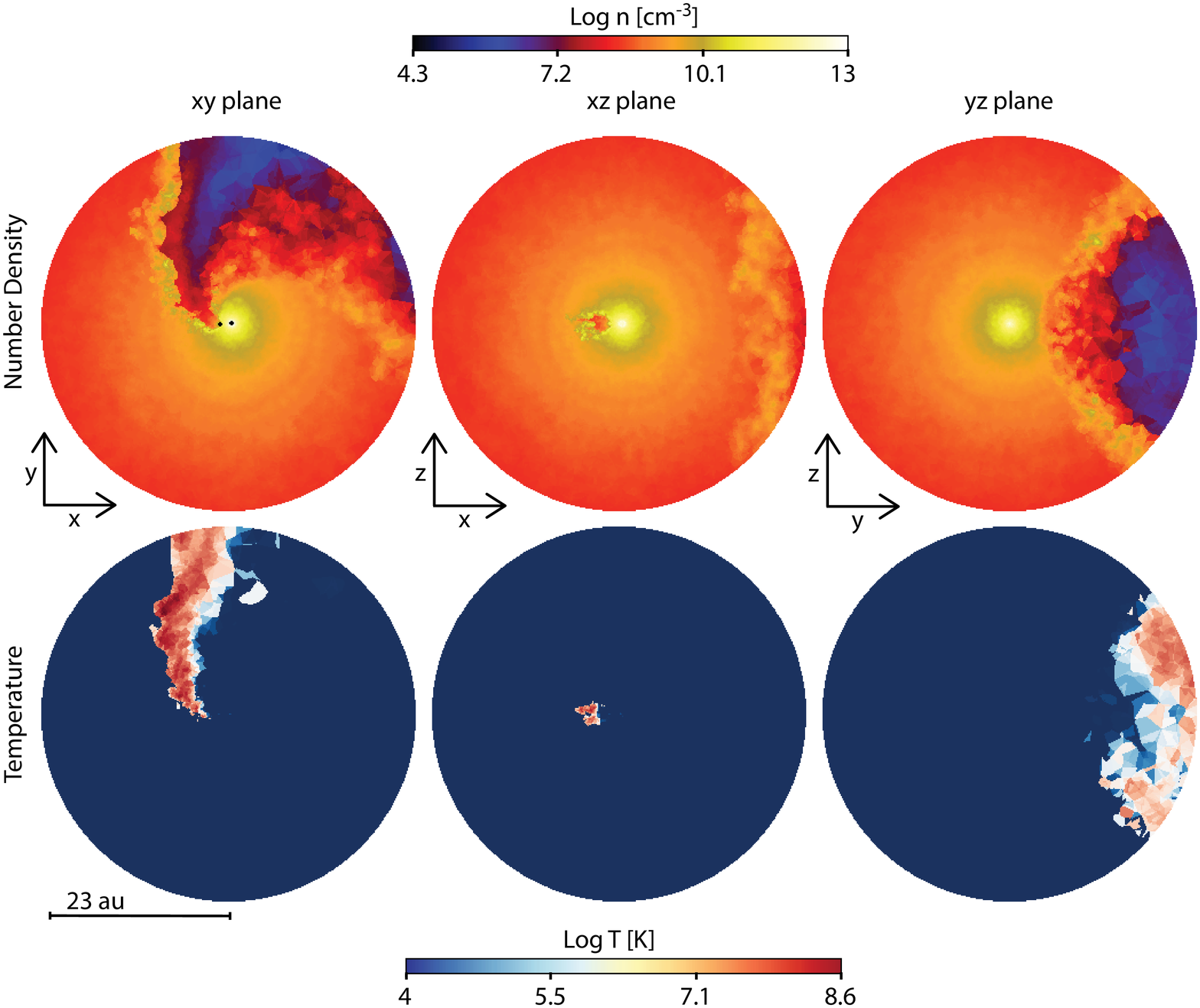}}
    \caption{Same as Fig.~\ref{fig:CaseA_small_dens_temp}, but for the Case~B simulation.}\label{fig:CaseB_small_dens_temp}
  \end{center}
\end{figure*}

To help the reader better understand the RT results, we first briefly describe the density and temperature structure in the $xy$, $xz$, and $yz$ planes. For more details on the hydrodynamical simulation results, see \citetalias{Madura_etA_2013}. Figs.~\ref{fig:CaseA_medium_dens_temp} and \ref{fig:CaseA_small_dens_temp} show the number density (top row) and temperature (bottom row) for the 155~au and 23~au Case~A simulations, respectively (the green circles in Fig.~\ref{fig:CaseA_medium_dens_temp} mark the domain size of the 23~au simulations). As described in \citetalias{Madura_etA_2013}, approaching periastron, the orbital speed of \etaA relative to \etaB increases to $\approx$~360~\kms, close to \etaA's wind speed (420~\kms), highly distorting the WWIR. The small stellar separation at periastron ($\sim$~1.5~au) also prevents \etaB's wind from reaching its terminal speed before colliding with \etaA's wind. Combined with the effects of radiative inhibition by \etaA \citepalias{Madura_etA_2013}, this leads to a significant slowing of \etaB's pre-shock wind, which alters the wind momentum ratio at the WWIR. This decreases the WWIR opening angle and moves the WWIR apex closer to \etaB. This behaviour occurs in the Case~B simulations in Figs.~\ref{fig:CaseB_medium_dens_temp} and \ref{fig:CaseB_small_dens_temp} as well, although lowering \MdotA further increases the WWIR opening angle and moves the WWIR apex closer to \etaA.

The above effects create a high asymmetry between the trailing and leading arms of the WWIR at periastron, as \etaB becomes deeply embedded in \etaA's dense inner wind. The only direction \etaB can effectively drive its wind is away from \etaA (top left panel of Figs.~\ref{fig:CaseA_small_dens_temp} and \ref{fig:CaseB_small_dens_temp}). This has a strong impact on the location of the shock-heated gas. As shown in the bottom left panel of Figs.~\ref{fig:CaseA_small_dens_temp} and \ref{fig:CaseB_small_dens_temp}, the post-shock secondary wind is heated to temperatures higher than \ten{6}~K only where the \etaB wind collides with the highly distorted leading arm of the WWIR (in red). The trailing wind of \etaB is unable to collide with \etaA's downstream wind, and so there is no hot shocked gas there. This effect is also visible in the bottom left panel of Figs.~\ref{fig:CaseA_medium_dens_temp} and \ref{fig:CaseB_medium_dens_temp}. However, in the larger domain simulations, some residual hot gas remains at the outermost right edge of the simulation domain in the $xy$ and $xz$ plane slices. This is the expanding, adiabatically-cooling remnant of the trailing arm of the WWIR from just before periastron.

Looking at the $xz$ plane (middle column of Figs.~\ref{fig:CaseA_medium_dens_temp}--\ref{fig:CaseB_small_dens_temp}), during periastron, \etaB passes behind \etaA, allowing, for a short period, the primary wind to expand in the apastron ($+x$) direction. During this time, the apex of the low-density cavity on the apastron side of the system fills with dense primary wind. The thickness and density of this inner primary wind region increases with \MdotA, since higher \MdotA move the apex of the WWIR farther from \etaA, allowing the primary wind to fill a larger volume in the apastron direction. Slices in the $xz$ plane also sample mostly cold (\ten{4}~K) gas from both winds, although the \etaB wind cavity in the Case~B simulations contains warmer material due to the larger WWIR opening angle and less oblique shocks \citepalias{Madura_etA_2013}.

Slices in the $yz$ plane (right-hand column of Figs.~\ref{fig:CaseA_medium_dens_temp}--\ref{fig:CaseB_small_dens_temp}) simply show that at periastron, the leading arm of the WWIR has passed through the $yz$ plane. There are now clear differences in both the density and temperature along the $y$ axis. The $-y$ side of the system still consists of unshocked \etaA wind, but to the $+y$ side there is a small cavity filled with hot, shocked \etaB wind that is bordered by dense post-shock \etaA wind. This cavity is larger in the Case~B simulations due to that simulation's larger WWIR opening angle.


\subsection{He ionization structure and influence of \MdotA}

\begin{figure*}
  \begin{center}
   \rmf{\includegraphics[width=174mm]{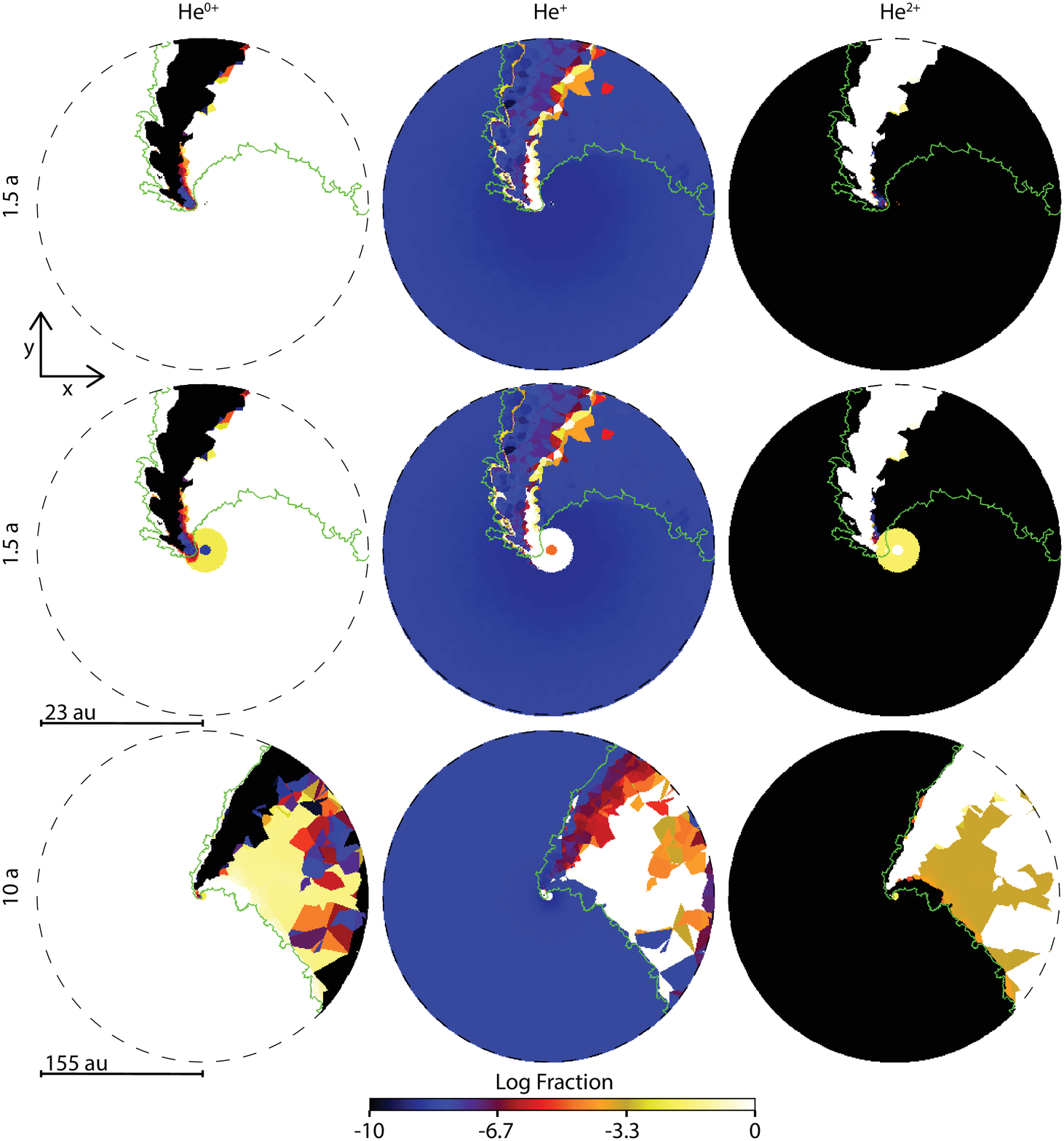}}
    \caption{Slices in the orbital plane through the 3D \SimpleX simulation volume for the Case~A simulation. Columns, from left to right, show the computed fractions of \ion{He}{0+}, \ion{He}{+}, and \ion{He}{2+} (log scale). Rows show, from top to bottom, the smaller (23~au) simulation domain without the inner ionization volumes due to the primary, the smaller (23~au) simulation domain with the \ion{He}{+} and \ion{He}{2+} inner ionization volumes due to the primary included, and the bigger (155~au) simulation domain including the \ion{He}{+} ionization volume due to the primary. In this and Fig.~\ref{fig:CaseB_xy}, the green line marks the edge of the pre-shock primary wind.}\label{fig:CaseA_xy}
  \end{center}
\end{figure*}

\begin{figure*}
  \begin{center}
   \rmf{\includegraphics[width=174mm]{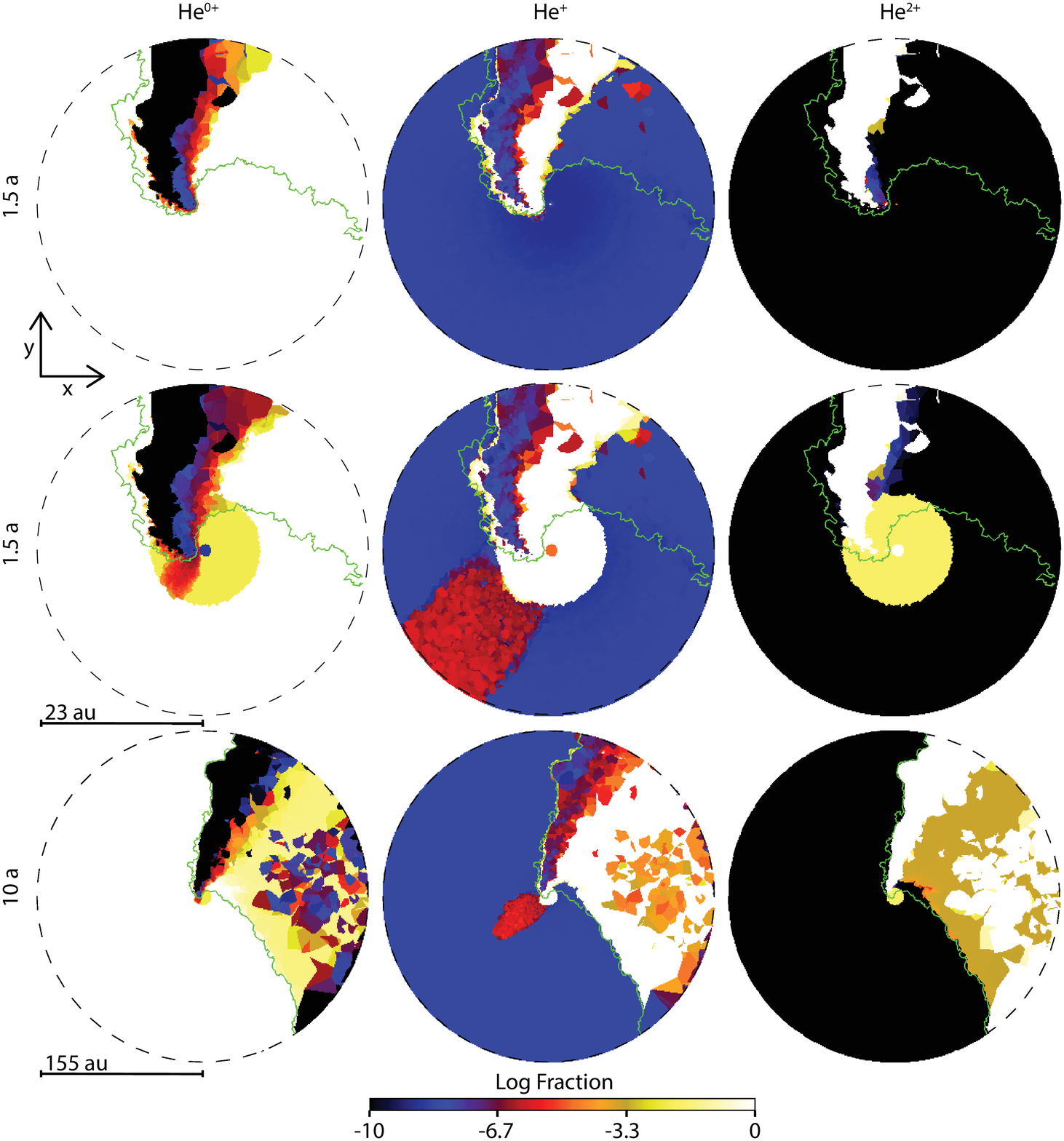}}
    \caption{Same as Fig.~\ref{fig:CaseA_xy}, but for Case~B.}\label{fig:CaseB_xy}
  \end{center}
\end{figure*}

Figs.~\ref{fig:CaseA_xy} and \ref{fig:CaseB_xy} present the He ionization structure in the orbital $xy$ plane for Cases~A and B, respectively. In all figures, columns illustrate, left to right, the fractions of \ion{He}{0+}, \ion{He}{+}, and \ion{He}{2+}. Rows display the 23~au domain without the inner ionization volume due to \etaA (top), the same domain with the inner \ion{He}{+} and \ion{He}{2+} ionization volumes (middle, filled yellow and blue circles in the left-most panel, respectively), and the larger (155~au) simulation domain with the inner \ion{He}{+} ionization volume due to \etaA included (bottom). The green contour highlights the boundary between the unperturbed pre-shock \etaA wind and the density-enhanced post-shock \etaA wind. Due to the complex situation at periastron, the green contour is meant more as a guide to better understand the results than a precise position for the pre-shock \etaA wind.

As discussed in \citetalias{Clementel_etA_2014b}, \etaB should not have many \ion{He}{+}-ionizing photons and, therefore, \ion{He}{2+} is produced principally through collisional ionization. As expected, neglecting the inner \ion{He}{2+} region in \etaA's wind, \ion{He}{2+} is only present in the hottest ($T \gtrsim \ten{5}$~K) regions of the system, i.e. the shock-heated gas in the leading arm of the WWIR, and the adiabatically-cooling hot gas in the remnant of the trailing arm (from before periastron passage).

Similar to the case at apastron \citepalias{Clementel_etA_2014b}, \etaB ionizing photons create different \ion{He}{+} structures depending on the value of \MdotA. In Case~A, the \ion{He}{0+}-ionizing photons are only able to ionize a thin layer within the leading arm of the WWIR, and a rather small volume of the diffuse \etaA wind that is expanding into the \etaB wind cavity located near the remnant of the WWIR's trailing arm (first row of Fig.~\ref{fig:CaseA_xy}). The dense primary wind expanding into the \etaB wind cavity is able to stop \etaB's \ion{He}{0+}-ionizing photons. The \ion{He}{+} present in the cooler, unshocked \etaB wind on the apastron side of the system is the previously photoionized gas that is slowly recombining. The inclusion of the inner \ion{He}{+} and \ion{He}{2+} regions due to \etaA have basically no effect on the larger-scale He ionization structures in Case~A (compare the top and middle rows of Fig.~\ref{fig:CaseA_xy}). The only noticeable difference is that the \ion{He}{0+}-ionizing photons are able to penetrate through the leading arm of the WWIR and further ionize a very small area inside the inner 3~au \etaA~\ion{He}{+} ionization volume (the red within the yellow circle, close to the centre in the $-y$ direction in the first panel of the middle row). For Case~A, the dense outer primary wind (i.e. the area beyond the green contour) is almost entirely \ion{He}{0+}.

The larger WWIR opening angle in Case~B, due to the lower \MdotA, allows the \ion{He}{0+}-ionizing photons to escape the inner core and singly-ionize He in a portion of the the expanding primary and secondary winds on the apastron side of the system (right-hand side of the panels in Fig.~\ref{fig:CaseB_xy}). Contrary to Case~A, introducing the regions of ionized He in \etaA's inner wind changes the extent of the ionization front in both the primary and secondary winds. The larger (7.5~au) \ion{He}{+} zone in Case~B allows ionizing photons to penetrate deeper into the receding primary wind, producing a region of \ion{He}{+} (although the fraction of \ion{He}{+} remains quite low, see the red region in the middle column in the middle and bottom rows of Fig.~\ref{fig:CaseB_xy}). The \ion{He}{+} on the apastron side of the system is a combination of ongoing photoionization and previously ionized gas which is slowly recombining. However, as in Case~A, the dense primary wind is mostly composed of \ion{He}{0+} and very effectively traps \etaB's \ion{He}{0+}-ionizing photons (bottom row of Fig.~\ref{fig:CaseB_xy}).


\section{Discussion}\label{sec:Discussion}


From the discussions in Sections~\ref{sec:Method} and \ref{sec:Results}, it is evident that the interpretation of the results at periastron is much more difficult then at apastron, especially in terms of trying to understand how the results relate to observations of the various He line profiles \citep[][]{Nielsen_etA_2007, Damineli_etA_2008b}. Nevertheless, we can use our results to try to better constrain the regions where the He emission and absorption features are generated. We focus on the Case~A simulation results, since as discussed in \citetalias{Madura_etA_2013, Clementel_etA_2014b}, and above, they most likely better represent the current value of \MdotA. We note that spectroscopic phase, defined by e.g. the disappearance of the narrow component of the \HeI lines (\citealt{Damineli_etA_2008b}, or alternatively, the minimum in the observed X-ray light curve, \citealt{Corcoran_etA_2010}), may differ from true orbital phase by up to one month.

Our results show that, at periastron, \etaB is able to singly-ionize He in only a very thin layer of the post-shock \etaA wind along the WWIR leading arm, and a very small region close to the stars. The denser \etaA wind, expanding into the secondary wind cavity, is effectively able to stop the \ion{He}{0+}-ionizing photons. Therefore, the lower-density receding secondary wind on the apastron side and the adiabatically-cooling trailing arm of the WWIR from before periastron are shielded from ionizing photons from \etaB. We do not expect this material to be neutral due to the long recombination time-scales for gas at these densities ($n < \ten{7}$~\per{cm}{-3}) and temperatures ($\gtrsim$~\ten{4}~K). The disappearance of the shock-heated region in the trailing arm of the WWIR, caused by the curvature of the WWIR, causes a strong asymmetry in the high temperature gas which is now mainly present in the WWIR leading arm. The He in this very hot region ($T > \ten{6}$~K) is collisionally doubly-ionized. With the exception of the inner \ion{He}{+} and \ion{He}{2+} ionization regions due to \etaA's ionizing flux, He in the extended primary wind is mostly neutral. This includes post-shock gas in the remnant of the trailing arm, and a significant portion of the extended leading arm, of the WWIR.

Observations show that the broad \eal{He}{I} lines are blueshifted for most of the orbital period, both in emission and absorption. The P~Cygni absorption starts increasing approximately three weeks before spectroscopic phase zero, reaching maximum absorption around phase zero, before decreasing to a complete disappearance $\sim$20~days after phase zero \citep{Nielsen_etA_2007, Damineli_etA_2008a}. The radial velocity of the absorption reaches its maximum blueshifted value ($-610$~\kms) shortly before periastron, and then suddenly shifts during periastron down to $\sim -250 \kms$. This absorption must be produced by material between the observer and \etaA, either in the WWIR or in the pre-shock \etaA wind.

Because the radial velocity variations seen in the \eal{He}{I} $\lambda$7067 absorption are reminiscent of those observed from a star in a highly eccentric orbit, \citet{Nielsen_etA_2007} assumed that the \eal{He}{I} absorption is associated with the pre-shock wind of \etaA, ionized by the far-UV flux from \etaB. However, this assumption leads to an unusual, extreme system mass ratio of $q \sim 0.1$, with a mass for \etaB of 210~\Ms and a mass for \etaA of $\sim 20$~\Ms, in strong disagreement with the most commonly accepted values. In contrast, based on the maximum velocity of the absorption seen in the \eal{He}{I} line and the requirements of high excitation flux and gas density, \citet{Damineli_etA_2008a} conclude that the only plausible locations for the formation of the broad \eal{He}{I} absorptions are the walls of the WWIR. Our RT results favour this second scenario. At periastron, \etaB is deep within the primary wind and its \ion{He}{0+}-ionizing photons are not able to reach the pre-shock \etaA wind or penetrate beyond the dense layer of post-shock \etaA wind. As shown in Fig.~\ref{fig:CaseA_xy} (central column, middle and bottom rows), He in the pre-shock \etaA wind is almost entirely neutral. The region of \ion{He}{+} is confined to the centre of the system.

Modeling of \ec's X-ray light curve and observed broad, extended emission from numerous forbidden emission lines have helped constrain the orbital inclination ($i$), argument of periapsis ($\omega$), and sky position angle (PA$_{z}$) of the \ec binary system \citep{Okazaki_etA_2008, Parkin_etA_2009, Madura_etA_2012}. It is generally agreed that the \ec binary is inclined away from the observer by roughly $\sim 45^{\circ}$, with $\omega \approx 240^{\circ}$--$285^{\circ}$ and PA$_{z} \approx 302^{\circ}$--$327^{\circ}$, which places \etaB at apastron on the observer's side of the system and implies that \etaB orbits clockwise on the sky. With this orientation of the binary on the sky, at apastron, our line-of-sight (LOS) is more perpendicular to the apex of the WWIR surface. Based on the results in \citetalias{Clementel_etA_2014b}, we propose that the much denser WWIR (specifically, the compressed post-shock \etaA wind) should dominate the P~Cygni absorption at phases around apastron. However, because our LOS is more perpendicular to the WWIR around apastron, the column density of \ion{He}{+} material between us and the continuum source \etaA is relatively modest, resulting in a moderate amount of absorption. Moreover, because we are viewing the WWIR inclined away from us at roughly 45~degrees, material flowing toward us within the wall of post-shock \etaA wind will have a maximum LOS velocity of $\sim$~300~\kms, since the maximum true velocity of the gas within the compressed wall of post-shock \etaA wind will be roughly equal to the terminal velocity of \etaA's wind.

When approaching periastron, the trailing arm of the WWIR sweeps across our LOS. When this occurs, our LOS is parallel to and intersecting the WWIR. This results in a significantly increased column of \ion{He}{+} between us and \etaA, producing an increase in the amount of \eal{He}{I} absorption. Because we are now looking directly down the WWIR, we are also seeing the fastest material in LOS, resulting not only in an increase in the amount of \eal{He}{I} absorption, but also an increase in the blue-shifted velocity of this absorption. This scenario is in qualitative agreement with both the increase in the absorption and its shift toward more blue-shifted velocities of the observed \eal{He}{I} P~Cygni profiles going into periastron. Just after periastron, the strong \eal{He}{I} $\lambda$7067 absorption component rapidly vanishes. This is also consistent with our results, since we see that at periastron, \etaB is no longer able to ionize the WWIR or the primary wind. With the observer on the apastron side of the system, the gas in LOS, further away from the central region, cannot be ionized by \etaB, resulting in effectively no \ion{He}{+} between us and the central \etaA~\ion{He}{+} region, and thus a lack of \eal{He}{I} absorption.

The results in \citetalias{Clementel_etA_2014b} imply that the \eal{He}{I} emission seen around apastron should arise from a combination of the inner \etaA~\ion{He}{+} zone, the pre-shock \etaA wind close to the WWIR apex ionized by \etaB, and the layer of post-shock \etaA wind in the WWIR that is also ionized by \etaB. The different regions, depending on the orbital phase, where the \eal{He}{I} emission can form might help explain the multiple emission components observed in the line profiles \citep{Nielsen_etA_2007}. At periastron, as discussed in \citetalias{Madura_etA_2013}, the WWIR apex is inside \etaA's \ion{He}{+} region. The fact that \etaB is able to ionize an extremely small region close to the centre of the system during periastron means that the dominant source of \eal{He}{I} emission at periastron is the inner \etaA~\ion{He}{+} region. Based on our models, little to no \eal{He}{I} emission is expected from the WWIR at and just after periastron.

Just before periastron, the \eal{He}{II} $\lambda$4686 line intensity increases suddenly and then drops sharply to zero, after which it recovers to a second peak before declining back to zero \citep{Steiner_Daminelli_2004, Martin_etA_2006, Mehner_etA_2011, Teodoro_etA_2012}. Based on the density and energy required for the formation of this line, the most plausible region in which it can form during periastron passage is close to the WWIR apex \citep{Martin_etA_2006, Teodoro_etA_2012}. Our results support the scenario proposed by \citetalias{Madura_etA_2013} in which the \eal{He}{II} $\lambda$4686 emission, observed at phases $\phi \sim$~0.98--1.3, is generated in the inner \ion{He}{+} region of \etaA's dense wind, with the necessary \ion{He}{+}-ionizing photons coming from the radiatively-cooling shocks in the WWIR. Our RT results show that, near periastron, \etaB's ionizing flux is unable to produce a significant, extended region of \ion{He}{+} in either the dense pre-shock \etaA wind or the densest parts of the WWIR. Even though the unshocked \etaB wind remains in the \ion{He}{+} state during periastron, as discussed by e.g. \citet{Martin_etA_2006, Teodoro_etA_2012}, the much lower-density \etaB wind is incapable of adequately explaining the observed \eal{He}{II} $\lambda$4686 emission around periastron. Thus, the only remaining region that can give rise to the observed \eal{He}{II} $\lambda$4686 emission is the \ion{He}{+} zone in \etaA's dense inner wind.

The sharp drop in \eal{He}{II} $\lambda$4686 emission seen near periastron might be caused by a combination of two effects. First, we note that at periastron, the WWIR opening angle decreases considerably due to the much lower wind momentum ratio, caused by a large decrease in the pre-shock velocity of \etaB's wind due to the decreased orbital separation and radiative inhibition effects (see Section~\ref{sec:Results} and \citetalias{Madura_etA_2013}). At periastron, \etaB and the narrowed WWIR are also behind \etaA and its optically-thick wind (relative to the observer). Thus, \etaB and the WWIR are both `eclipsed' by \etaA and its dense wind at periastron. Second, if the WWIR `collapses' at periastron due to the reduced \etaB wind speed (see discussions in e.g. \citealt{Corcoran_etA_2010,Parkin_etA_2011}; \citetalias{Madura_etA_2013}), the radiatively-cooling WWIR shocks might disappear. If this occurs, the source of \ion{He}{+}-ionizing photons (i.e. the soft X-rays generated in the radiative shocks) necessary to produce the \eal{He}{II} emission might also disappear. There would then be no significant \eal{He}{II} emission at periastron. We believe that a combination of a wind-eclipse by \etaA and a `collapse' of the WWIR is currently the best scenario for explaining the observed drop in \eal{He}{II} $\lambda$4686 emission near periastron.

The Case~B simulation results are similar to those of Case~A, but with two major differences. The 1D \textsc{cmfgen} models \citep{Hillier_etA_2001, Hillier_etA_2006} predict a radius for the inner \ion{He}{+} region, due to \etaA ionization, of $\sim 7$~au, more than a factor of two larger than in Case~A. Even without this larger ionization region, \etaB photons are able to ionize the denser primary wind that is expanding into the \etaB wind cavity and reach the receding unshocked \etaB wind on the apastron side of the system. Qualitatively, we expect these differences to produce variations, compared to Case~A, both in the strength and velocity of the He emission and absorption lines. Moreover, the fact that \etaB ionizing photons are still able at apastron to reach the periastron side of the system, have important implications on the phases at which these observed features occur. The increase in WWIR opening angle as \MdotA drops should also alter the timing of when specific He emission and absorption features are observed.

Finally, we wish to comment on one detail not implemented in our 3D simulations that may alter slightly the He ionization structure around periastron. Our current models do not include an ionizing source at the location of \etaA. The reasons for this, discussed in \citet{Clementel_etA_2014a}, are related to the complexities of proper spectroscopic modeling of the dense, extended wind photosphere of \etaA, which requires inclusion of the many ionization states of metals up to and including iron. However, we note that during periastron passage, \etaB and the WWIR penetrate deep within the extended wind of \etaA, down into its hotter regions where He is ionized to \ion{He}{+}. This `bore-hole' effect has many repercussions for multiwavelength observables (see \citealt{Madura_Owocki_2010, Madura_2012, Groh_etA_2012a, Groh_etA_2012b}), and it may affect the He ionization structure of the innermost regions of the \ec system as well. This is because the hole created by the WWIR cavity within the innermost \etaA wind may provide an escape path for higher-energy photons from \etaA. These higher energy \etaA photons may be able to escape through the WWIR cavity and help ionize He in portions of the dense arms of the WWIR, and possibly a very thin layer of pre-shock \etaA wind located beyond the WWIR. Such an effect would be limited to around periastron and extremely short-lived since the WWIR geometry changes so rapidly. Escaping higher-energy photons from the interior of \etaA would only be able to escape in very specific directions and travel a limited distance before the WWIR geometry shifted and shut-off (or changed the direction of) the escaping radiation. Thus, the effect would be limited to the innermost regions. It is also questionable whether such changes to the He ionization structure would even be observable, since the WWIR and bore-hole would be located on the opposite side of \etaA and directed away from the observer. More detailed 3D radiation-hydrodynamics simulations are necessary to accurately determine how much this bore-hole effect may alter the inner He ionization structure and affect observations of different He lines.


\section{Summary and Conclusion}\label{sec:Summary}

We presented the ionization structure of helium at periastron in the inner regions of the \ec binary system due to the hot secondary star's ionizing flux and collisional ionization. In our analysis, we considered two different \MdotA, i.e. Case~A ($\MdotA = \sci{8.5}{-4}$~\Msy) and Case~B ($\MdotA = \sci{4.8}{-4}$~\Msy). We created He ionization maps by post-processing 3D SPH simulations of the \ec system with \SimpleX. Below we summarize our most important results and conclusions.

\begin{enumerate}[leftmargin=*, label=\arabic*.]

\item At periastron, in the Case~A simulation, the dense \etaA wind expanding into the \etaB wind cavity is able to stop \etaB's \ion{He}{0+}-ionizing photons. Helium is only singly-ionized by \etaB in a very thin layer along the WWIR leading arm, and in a very small region close to the stars.

\item In Case~B, the larger WWIR opening angle and lower \MdotA allow \etaB to ionize a region of the lower-density unshocked \etaB wind expanding in the apastron direction.

\item The inner \ion{He}{+} volume due to \etaA does not produce relevant differences in the final ionization structures due to \etaB in simulation Case~A. In Case~B, the larger \etaA\ \ion{He}{+} radius allows the \etaB ionizing photons to reach deeper into the dense \etaA wind and create a lower ionization fraction in the pre-shock primary wind.

\item Collisional ionization creates \ion{He}{2+} in the hot shock-heated gas in the leading arm of the WWIR and in the adiabatically-cooling remnant of the trailing arm formed just before periastron.

\item Our RT results support a binary orientation in which apastron is on the observer's side of the system, with the trailing arm of the WWIR sweeping across our LOS as periastron is approached.

\item The small \ion{He}{+} region at the centre of the system is the main region were \eal{He}{I} emission can form during periastron. There is little to no \eal{He}{I} absorption at periastron because \etaB's ionizing flux is effectively trapped by \etaA's dense wind, leading to a lack of dense \ion{He}{+} gas between us and the continuum source \etaA. This is consistent with the observed changes in strength and radial velocity of the \eal{He}{I} P~Cygni lines across periastron.

\item A lack of any significant \ion{He}{+} in/near the WWIR at periastron, due to \etaB, implies that the increased \eal{He}{II} $\lambda$4686 emission that is observed around periastron must arise from the inner \ion{He}{+} region in \etaA's dense wind. Our results favour a scenario in which a combination of a wind-eclipse by \etaA and a `collapse' of the WWIR are responsible for the observed drop in \eal{He}{II} $\lambda$4686 emission at periastron.

\end{enumerate}

Our model He-ionization structures at periastron are in qualitative agreement with available observations of \ec's He lines. In the future, we plan to generate synthetic He line profiles for comparison to available observational data in order to place tighter constraints on the orbital, stellar, and wind parameters of \ec. These synthetic spectra, together with those generated for the spectroscopic high state around apastron, will help us better understand the numerous observed spectral features generated in the inner $\sim 150$~au of the system. Our final aim is to use 3D time-dependent radiation-hydrodynamics simulations to more properly follow the ionization and recombination of the gas without postprocessing. This is particularly important in the rapidly changing conditions and highly asymmetric situation at periastron.


\section*{Acknowledgments}

T. I. M. is supported by an appointment to the NASA Postdoctoral Program at the Goddard Space Flight Center, administered by Oak Ridge Associated Universities through a contract with NASA.


\ifdraft{\bibliography{biblio}} {\bibliography{biblio}}

\label{lastpage}

\end{document}